\documentclass[12pt]{article}
\usepackage{latexsym} \usepackage{amsfonts}
\begin{document}

\begin{center}
{\Large \bf  Massive Ghosts in Softly Broken SUSY Gauge\\[0.5cm]
Theories} \vspace{1cm}

{\large \bf D.I.Kazakov$^{\dag , \ \ddag}$ and
V.N.Velizhanin$^\dag$ }\vspace{0.7cm}

{\it $^\dag$ Bogoliubov Laboratory of Theoretical Physics, Joint
Institute for Nuclear Research, Dubna, Russia \\[0.2cm] and\\[0.2cm]
$^\ddag$ Institute for Theoretical and Experimental Physics,
Moscow, Russia}
\end{center}\vspace{0.5cm}

\begin{abstract}
It is shown that, due to soft supersymmetry breaking in gauge theories
within the superfield formalism, there appears the mass
for auxiliary gauge fields. It enters into the RG equations for
soft masses of physical scalar particles and can be eliminated by
solving its own RG equation. Explicit solutions up to the three-loop
order in the general case and in the MSSM are given. The arbitrariness
in choosing the initial condition is discussed.
\end{abstract}\vspace{0.3cm}

\section{Introduction}
Recently, it has been realized~\cite{Y,JJ9709,AKK} that
renormalizations in a softly broken SUSY theory are not
independent but follow from those of an unbroken SUSY theory.
According to the approach advocated in Refs.~\cite{AKK,K98}, one
can perform the renormalization of a softly broken SUSY theory in
the following straightforward way:

{\it One takes  renormalization constants of a rigid theory,
calculated in some massless scheme, substitutes for the
rigid couplings (gauge and Yukawa) their modified expressions,
that depend on a Grassmannian variable, and expands over this
variable.}

This gives renormalization constants for the soft terms.
Differentiating them with respect to a scale, one can find
corresponding renormalization-group equations.

Thus, the soft-term renormalizations are not independent but can be calculated
from the known renormalizations of a rigid theory with the help of the
differential operators. Explicit form of these operators has been found in a
general case and in some particular models like SUSY GUTs or the
MSSM~\cite{AKK}. The same expressions have been obtained also in a somewhat
different approach in Ref.~\cite{JJ9709,JJ9712,GR}.

There is, however, some minor difference. The authors of~\cite{JJ9709,JJ9712}
have used the component approach, while in~\cite{Y,AKK,K98}, the superfield
formalism is exploited. This creates the usual difference in gauge-fixing and
ghost field terms and in the renormalization scheme. The latter is related to
the choice of regularization. In~\cite{JJ9709,JJ9712}, the dimensional
reduction (DRED) regularization is used. In this case, one is bounded to
introduce the so-called $\epsilon$-scalars to compensate the lack of bosonic
degrees of freedom in 4-2$\epsilon$ dimensions. These $\epsilon$-scalars in due
course of renormalization acquire a soft mass that enters into the RG equations
for soft masses of physical scalar particles. This problem has been discussed
in~\cite{JJMVY}. If one gets rid of the $\epsilon$-scalar mass by changing the
renormalization scheme, ${\mathrm {DRED}}\rightarrow{\mathrm {DRED'}}$, there
appears an additional term in RG equations for the soft scalar
masses~\cite{MV,JJ9405} called X~\cite{JJ9712}. This term is absent in RG
equations in Refs.~\cite{Y,AKK,K98}.

We have to admit that, indeed in our approach, though the $\epsilon$-scalars in
the superfield formalism are absent, that term appears in higher orders and is
related to the soft masses of other unphysical particles, the auxiliary gauge
fields. We show below how it emerges in the superfield formalism and coincides
with that of Ref.~\cite{JJ9712}. Thus, the two approaches finally merge.

\section{Massive Auxiliary Fields}

Consider an arbitrary $N=1$ SUSY gauge theory with unbroken SUSY
within  the superfield formalism. The Lagrangian of a rigid theory
is given by
\begin{eqnarray}
{\cal L}_{rigid} &=& \int d^2\theta~\frac{1}{4g^2}{\rm Tr}W^{\alpha}W_{\alpha}
+ \int d^2\bar{\theta}~\frac{1}{4g^2}{\rm Tr} \bar{W}_{\dot
\alpha}\bar{W}^{\dot \alpha} \label{rigidlag} \\ &+& \int d^2\theta
d^2\bar{\theta} ~~\bar{\Phi}^i (e^{V})^j_i\Phi_j + \int
 d^2\theta ~~{\cal W} + \int d^2\bar{\theta} ~~\bar{\cal W},  \nonumber
\end{eqnarray}
where
 $$W_{\alpha}=-\frac 14\bar D^2e^{-V}D_\alpha e^V,
\ \ \ \bar W_{\dot \alpha}=-\frac 14 D^2e^{-V}\bar D_{\dot \alpha} e^V,$$ are
the gauge field strength tensors and the superpotential ${\cal W}$ has the form
\begin{equation}
{\cal  W}=\frac{1}{6}y^{ijk}\Phi_i\Phi_j\Phi_k +\frac{1}{2}
M^{ij}\Phi_i\Phi_j.\label{rigidsuppot}
\end{equation}
To fix the gauge, the usual gauge-fixing term can be introduced. It is useful
to choose it in the form
\begin{equation}\label{gf}
{\cal L}_{gauge-fixing}= -~\frac{1}{16}\int d^2\theta  d^2\bar{\theta}
{\rm Tr}\left(\bar{f}f + f\bar{f}\right)
\end{equation}
where  the gauge fixing condition is taken as
\begin{eqnarray}
f=\bar{D}^2\frac{V}{\sqrt{\xi g^2}}\;,
 ~~~ \bar f = D^2\frac{V}{\sqrt{\xi g^2}}\; . \label{gcond}
\end{eqnarray}
 Then, the corresponding ghost term is \cite{FP}
\begin{equation}\label{ghost}
  {\cal L}_{ghost}=i\int d^2\theta~\frac{1}{4}{\rm Tr}~b\,
\delta_{c}f -i\int d^2\bar{\theta}~\frac{1}{4}{\rm
Tr}~\bar{b}\, \delta_{\bar c}\bar f,
\end{equation}
where $c$ and $b$ are the Faddeev--Popov ghost and antighost
chiral superfields, respectively, and $\delta_{c}$ is the
gauge transformation with the replacement of gauge superfield
parameters $\Lambda (\bar \Lambda)$ by chiral (antichiral) ghost
fields $c (\bar c)$.

For our choice of the gauge-fixing condition,
%after rescaling $c \to c/g$
the gauge transformation of $f$ looks like
\begin{equation}\label{trans}
\delta_{\Lambda} f = \bar D^2\delta_{\Lambda} \frac{V}{\sqrt{\xi g^2}}
= i\bar D^2 \frac{1}{\sqrt{\xi g^2}}{\cal L}_{V/2}[\Lambda+\bar
\Lambda + \coth({\cal L}_{V/2})(\Lambda - \bar \Lambda)],
\end{equation}
where ${\cal L}_{X}Y\equiv[X,Y]$. Equation (\ref{ghost}) then takes the form
\begin{eqnarray}
{\cal L}_{ghost}&=&- \int d^2\theta~\frac{1}{4}{\rm
Tr}~b\bar{D}^2\frac{1}{\sqrt{\xi g^2}}{\cal L}_{V/2}[c+\bar c +
\coth({\cal L}_{V/2})(c - \bar c)] + \hspace{0.3cm} h.c.
\nonumber\\ &=& \int d^2\theta d^2\bar{\theta}~{\rm Tr}~
\left(\frac{b+ \overline{b}}{\sqrt{\xi g^2}}\right) {\cal
L}_{V/2}[c+\bar c + \coth({\cal L}_{V/2})(c - \bar c)]\label{gh}\\
&&\hspace*{-1.5cm}=\int d^2\theta d^2\bar{\theta}~{\rm Tr}~
\left(\frac{b+ \overline{b}}{\sqrt{\xi g^2}}\right) \left(\left
(c-\overline{c}\right) + \frac{1}{2} \Big[V,
\left(c+\overline{c}\right)\Big]+ \frac{1}{12} \bigg[V, \Big[V,
\left(c-\overline{c}\right)\Big]\bigg]+ ... \right). \nonumber
\end{eqnarray}

The resulting Lagrangian together with the gauge-fixing and the ghost terms are
invariant under the BRST transformations.      For a rigid theory in our
normalization of the  fields, they have the form~\cite{FP}
\begin{eqnarray}
&&\delta V=\epsilon {\cal L}_{V/2}[c+\bar c + \coth({\cal
L}_{V/2})(c - \bar c)], \nonumber \\
% &&\delta\,\Phi_i=-\epsilon c^aT^a_{ij}\Phi_j\ , \;\;\;\;\; \delta\, {\bar
%\Phi}_i=\epsilon {\bar \Phi}_j{\bar c}^aT^a_{ji}\ ,\nonumber \\
&&\delta\, c^a=-\frac {i}{2}\epsilon f^{abc}c^b c^c\ ,\;\;\;\;\;
\delta\, {\bar c}^a=-\frac {i}{2}\epsilon f^{abc}{\bar c}^b {\bar
c}^c\ ,\nonumber \\ &&\delta\, b^a = \frac{1}{8}\epsilon \bar
D^2 \bar f^a\ ,  \;\;\;\; \delta\, {\bar b}^a =
\frac{1}{8}\epsilon  D^2 f^a. \label{BRSTr}
\end{eqnarray}

To  perform the SUSY breaking, that satisfies the requirement of "softness",
one can introduce a gaugino mass term as well as cubic and quadratic
interactions of  scalar superpartners of the matter fields~\cite{spurion}
\begin{eqnarray}
-{\cal L}_{soft-breaking} &=&\left[ \frac{M}{2}\lambda\lambda
+\frac 16 A^{ijk} \phi_i\phi_j\phi_k+ \frac 12 B^{ij}\phi_i\phi_j
+h.c.\right] +(m^2)^i_j\phi^{*}_i\phi^j,\label{sofl}
\end{eqnarray}
where $\lambda$ is  the gaugino field, and $\phi_i$ is the lowest component of
the chiral matter superfield.

One can rewrite  the Lagrangian (\ref{sofl}) in terms of  N=1
superfields introducing  the external spurion
superfields~\cite{spurion} $\eta=\theta^2$ and $\bar
\eta=\bar{\theta}^2$, where $\theta$ and $\bar \theta$ are
Grassmannian parameters, as~\cite{Y}
 \begin{eqnarray} {\cal
L}_{soft} &=& \int d^2\theta~\frac{1}{4g^2}(1-2M\theta^2) {\rm
Tr}W^{\alpha}W_{\alpha} + \int
 d^2\bar{\theta}~\frac{1}{4g^2}(1-2\bar{M}\bar{\theta}^2) {\rm
Tr}\bar{W}^{\dot \alpha}\bar{W}_{\dot \alpha}  \nonumber \\
&&+\int d^2\theta d^2\bar{\theta} ~~\bar{\Phi}^i(\delta^k_i
-(m^2)^k_i\eta \bar{\eta})(e^V)^j_k\Phi_j   \label{sofl2} \\ &&+
\int  d^2\theta \left[\frac 16 (y^{ijk}-A^{ijk}
\eta)\Phi_i\Phi_j\Phi_k+ \frac 12 (M^{ij}-B^{ij}\eta )
\Phi_i\Phi_j \right] +h.c. \nonumber
\end{eqnarray}
Thus, one can interpret the soft terms as the modification of the
couplings of a rigid theory. The couplings become external
superfields depending on Grassmannian parameters $\theta$ and
${\bar \theta}$. To get the explicit expression for the modified
couplings, consider eqs.(\ref{sofl2}). The first two terms
give~\cite{AKK}
\begin{equation}
\frac{1}{g^2}\rightarrow\frac{1}{{\tilde g}^2}=\frac{1-M\theta ^2-{\bar M}{\bar
\theta}^2}{g^2}. \label{gtilold}
\end{equation}
Since the gauge field strength tensors $W_{\alpha}$ (${\bar
W}_{\alpha}$) are chiral (antichiral) superfields, they enter into
the chiral (antichiral) integrands in eq.(\ref{sofl2}),
respectively. Correspondingly, the $M\theta ^2$ term of
eq.(\ref{gtilold}) contributes to the chiral integral, while the
${\bar M} {\bar \theta}^2$ term contributes to the antichiral one.
There is no $\theta ^2 {\bar \theta}^2$ term in
eq.(\ref{gtilold}), since it is neither chiral, no antichiral and
gives no contribution to eq.(\ref{sofl2}).

However, modifying the gauge coupling in the gauge part of the Lagrangian, one
has to do the same in the gauge-fixing (\ref{gcond}) and ghost (\ref{gh}) parts
in order to preserve the BRST invariance. Here one has the integral over the
whole superspace rather than the chiral one. This means that if one adds to
eq.(\ref{gtilold}) a term proportional to $\theta ^2{\bar{\theta}}^2$, it gives
a nonzero contribution. Moreover, even if this term is not added, it reappears
as a result of renormalization.

We suggest the following modification of eq.(\ref{gtilold})
\begin{equation}
\frac{1}{g^2}\rightarrow\frac{1}{{\tilde g}^2}=\frac{1-M\theta
^2-{\bar M}{\bar \theta}^2-\Delta \theta ^2{\bar{\theta}}^2}{g^2},
\end{equation}
which  gives  the final expression for the soft  gauge coupling
\begin{eqnarray} \tilde{g}^2 = g^2\left(1 + M\theta^2 +
\bar M\bar{\theta}^2 + 2 M {\bar M}\theta^2 \bar{\theta}^2 +
\Delta \theta^2 \bar{\theta}^2\right).  \label{Tildeg}
\end{eqnarray}
In our previous papers~\cite{AKK,K98}, this $\Delta$ term was absent. It will
be clear below that it is self-consistent to put $\Delta=0$ in the lowest order
of perturbation theory, but it appears in higher orders due to
renormalizations.

One has to take into account, however,  that, since the gauge-fixing parameter
$\xi$ may be considered as an additional coupling, it also becomes an external
superfield and has to be modified. The soft expression can be written as
\begin{equation}\label{xi}
  \tilde \xi = \xi \left(1+ x\theta^2 + \bar x \bar \theta^2 + (x\bar x
  + z)\theta^2\bar \theta^2\right),
\end{equation}
where parameters $x$ and $z$ can be obtained by solving the corresponding RG
equation (see Appendix A).

Having this in mind, we perform the following modification of the gauge fixing
condition (\ref{gcond}) first used in~\cite{Kond}
\begin{equation}\label{gcond2}
  f \to \bar D^2\frac{V}{\sqrt{\tilde \xi {\tilde g}^2}}, \ \ \ \bar f
  \to D^2\frac{V}{\sqrt{\tilde \xi {\tilde g}^2}}.
\end{equation}
Then, the gauge-fixing term (\ref{gf}) becomes
\begin{equation}\label{gfm}
{\cal L}_{gauge-fixing}= -~\frac{1}{8}\int d^2\theta d^2\bar{\theta} ~{\rm
Tr}\left( \bar D^2\frac{V}{\sqrt{\tilde \xi {\tilde g}^2}}
D^2\frac{V}{\sqrt{\tilde \xi {\tilde g}^2}}\right),
\end{equation}
This leads to the corresponding  modification of the associated
ghost term (\ref{ghost})
\begin{eqnarray}
{\cal L}_{ghost} &=&\int d^2\theta d^2\bar{\theta}~{\rm
Tr}~\frac{1}{\sqrt{\tilde \xi \tilde{g}^2}} \left(b+\overline{b}\right){\cal
L}_{V/2}[c+\bar c + \coth({\cal L}_{V/2})(c - \bar c)] . \label{ghs}
\end{eqnarray}

To understand the meaning of the $\Delta$ term, consider the
quadratic part of the ghost Lagrangian (\ref{ghs})
\begin{eqnarray}
{\cal L}_{ghost}^{(2)} &= &\int d^2\theta d^2\bar{\theta}~{\rm
Tr}~\frac{1}{\sqrt{\xi{g}^2}} \left(1 -\frac 12 M\xi
\theta^2 - \frac 12 {\bar M}\xi \bar{\theta}^2  - \frac 12  \Delta \xi \theta^2
\bar{\theta}^2\right) (b+\overline{b}) \left
(c-\overline{c}\right)\nonumber \\ &=&\int d^2\theta
d^2\bar{\theta}~{\rm Tr}~\frac{1}{\sqrt{\xi{g}^2}} \left(1 -
\frac 12 \Delta \xi \theta^2
\bar{\theta}^2\right) (b+\overline{b}) \left
(c-\overline{c}\right)\label{ghs2} \\ &-&\frac 12\int d^2\theta~{\rm
Tr}~\frac{1}{\sqrt{\xi{g}^2}} M\xi  bc + \frac 12\int
d^2\bar \theta~{\rm Tr}~\frac{1}{\sqrt{\xi{g}^2}}
\bar M \xi\bar b\bar c ,\nonumber
\end{eqnarray}
where we have used the explicit form of $\tilde \xi$ given in Appendix A.

If one compares this expression with the usual Lagrangian for the matter fields
(\ref{sofl2}), one finds an obvious identification of the second line with the
soft scalar mass term and the third line with the mass term in a
superpotential. Thus, $\Delta$ plays the role of a soft mass providing the
splitting in the ghost supermultiplet.

The other place where the $\Delta$-term appears is the gauge-fixing term
(\ref{gfm}). Here it manifests itself as a soft mass of the auxiliary gauge
field, one of the scalar components of the gauge superfield $V$.

To see this, consider the gauge-fixing term (\ref{gfm}) in more detail.
Expanding the vector superfield $V(x,\theta,\bar \theta)$ in components
\begin{eqnarray}
V(x, \theta, \bar \theta) & = & {\mathbb C}(x) + i\theta \chi (x) -i\bar \theta \bar \chi
(x) +  \frac{i}{2} \theta \theta N(x) - \frac{i}{2} \bar
 \theta \bar \theta \bar N(x)
 -  \theta \sigma^{\mu} \bar \theta v_{\mu}(x)\label{p}\\ &  &\hspace*{-2cm}
  + i \theta \theta
\bar \theta [\bar \lambda (x) + \frac{i}{2}\bar \sigma^{\mu}
\partial _{\mu} \chi (x)]
  -  i\bar \theta \bar \theta \theta [\lambda + \frac{i}{2}
\sigma^{\mu} \partial_{\mu} \bar \chi (x)] + \frac{1}{2} \theta \theta \bar
\theta \bar \theta [D(x) - \frac{1}{2}\Box {\mathbb C}(x)].\nonumber
  \end{eqnarray}
and substituting it into eq.(\ref{gfm}) one finds
\begin{eqnarray}
  {\cal L}_{gauge-fixing}&=&\frac{1}{2\xi g^2}\left[-(D-\Box {\mathbb C} -\Delta \xi {\mathbb C} +\frac i2
M \xi \bar N -\frac i2 \bar M \xi N)^2 - (\partial ^\mu v_\mu)^2 \right. \nonumber \\
&+& \left. (\bar N -i \bar M \xi {\mathbb C})\Box
( N +i M \xi {\mathbb C})+  2i(\lambda + \frac 12 \bar M\xi \chi )\sigma^\mu\partial_\mu (\bar
\lambda + \frac 12 M\xi \bar \chi ) \right. \nonumber \\
&-&\left.
 2(\lambda + \frac 12 \bar M\xi \chi)\Box \chi - 2(\bar
\lambda + \frac 12 M\xi \bar \chi )\Box \bar \chi -2i\Box \chi\sigma^\mu \partial_\mu \bar \chi \right].
 \label{com}
\end{eqnarray}
One can see from eq.(\ref{com}) that the parameter $M$, besides being the
gaugino soft mass, plays the role of a mass of the auxiliary  field $\chi$,
while $\Delta$ is the soft mass of the auxiliary  fields $N$ and ${\mathbb C}$.
All these fields are unphysical degrees of freedom of the gauge superfield.
They are absent in the Wess-Zumino gauge, however, when the gauge-fixing
condition is chosen in supersymmetric form (\ref{gf}), this gauge is no longer
possible, and the auxiliary fields $\chi$, $N$, and ${\mathbb C}$ survive.
Thus, the extra $\Delta$ term is associated with unphysical, ghost, degrees of
freedom, just like in the component approach, one has the mass of unphysical
$\epsilon$-scalars. When we go down with energy, all massive fields decouple,
and we get the usual nonsupersymmetric Yang-Mills theory.

The $\Delta$-term is renormalized and obeys its own RG equation which can be
obtained from the corresponding expression for the gauge coupling via
Grassmannian expansion. In due course of renormalization, this term is mixing
with the soft masses of scalar superpartners and gives an additional term in RG
equations for the latter ($X$ term of Ref.~\cite{JJ9712} mentioned above).

At the end of this section, we would like to comment on the BRST invariance in
a softly broken SUSY theory. The BRST transformations (\ref{BRSTr}) due to our
choice of normalization of the gauge and ghost fields do not depend on the
gauge coupling. Hence, in a softly broken theory they remain unchanged. One can
easily check that, despite the substitution $g^2 \to \tilde g^2$ and $\xi \to
\tilde \xi$, the softly broken SUSY theory remains BRST invariant~\cite{Kond}.

\section{RG Equations for the Soft Parameters.}

Thus,  following the procedure described in  Refs~\cite{AKK,K98}, to get the RG
equations for the soft terms, one has to modify the gauge ($g^2_i$) and Yukawa
($y_{ijk}$) couplings replacing them by  external superfields :
\begin{eqnarray}
\tilde{g}^2_i&=&g^2_i(1+M_i \eta+{\bar M}_i \bar{\eta}+(2 M_i
{\bar M}_i +\Delta_i) \eta \bar{\eta}), \label{g}\\
\tilde{y}^{ijk}&=&y^{ijk}-A^{ijk}\eta +\frac 12 (y^{njk}(m^2)^i_n
+y^{ink}(m^2)^j_n+y^{ijn}(m^2)^k_n)\eta \bar \eta, \label{y1}\\
\tilde{\bar y}_{ijk}&=&\bar y_{ijk} - \bar A_{ijk} \bar{\eta}+
\frac 12 (y_{njk}(m^2)_i^n +y_{ink}(m^2)_j^n+y_{ijn}(m^2)_k^n)\eta
\bar \eta  . \label{y2}
\end{eqnarray}
Then, the $\beta$ functions of  RG equations for the soft masses of scalar
superpartners of the matter fields and  for the mass of the auxiliary gauge
field are given by~\cite{AKK}
\begin{eqnarray}
[\beta_{m^2}]^i_j=D_2\gamma^i_j ,\\ \beta_{\Sigma_{\alpha_i}} =D_2
\gamma_{\alpha_i}, \label{s}
\end{eqnarray}
where $\gamma^i_j$ and $\gamma_{\alpha_i}=\beta_{\alpha_i}/{\alpha_i}$ are the
anomalous dimensions of the matter fields and of the gauge coupling,
respectively, and we have introduced the notation
 $$\Sigma_{\alpha_i}=M_i{\bar M}_i+\Delta_i.$$
The modified expression for the operator $D_2$ is
\begin{eqnarray}
D_2&=& \bar{D}_1 D_1 + \Sigma_{\alpha_i} \alpha_i \frac{\displaystyle \partial
}{\displaystyle \partial \alpha_i}
+\frac{1}{2}(m^2)^a_n\left(y^{nbc}\frac{\displaystyle \partial }{\displaystyle
\partial y^{abc}} +y^{bnc}\frac{\displaystyle \partial
 }{\displaystyle \partial y^{bac}}+ y^{bcn}\frac{\displaystyle \partial
}{\displaystyle \partial y^{bca}}\right. \nonumber \\ && \left.
+y_{abc}\frac{\displaystyle \partial }{\displaystyle \partial y_{nbc}}+
y_{bac}\frac{\displaystyle \partial }{\displaystyle \partial y_{bnc}}+
y_{bca}\frac{\displaystyle \partial }{\displaystyle \partial
y_{bcn}}\right)\label{D2}.
\end{eqnarray}
It coincides now with that of Ref.~\cite{JJ9712} with $X_i=\Delta_i$.

To find $\Sigma_{\alpha_i}$, one can use  equation (\ref{s}). In particular,
using the expression for the anomalous dimension $\gamma_\alpha$ in case of a
single non-abelian gauge group calculated up to three loops \cite{JJ96061}
\begin{eqnarray}
\gamma_{\alpha}&=&\alpha Q+2\alpha ^2QC(G)-\frac{2}{r}\alpha
{\gamma^i_j}^{(1)}C(R)^j_i-\alpha ^3Q^2C(G)+4\alpha
^3QC^2(G)\nonumber\\ &&-\frac{6}{r}\alpha^3QC(R)^i_jC(R)^j_i
%&&-\frac{2}{r}\alpha {\gamma^i_j}^{(2)}C(R)^j_i
-\frac{4}{r}\alpha^2C(G) {\gamma^i_j}^{(1)}C(R)^j_i
+\frac{3}{r}\alpha y^{ikm}y_{jkn}
{\gamma^n_m}^{(1)}C(R)^j_i\nonumber\\ &&+\frac{1}{r}\alpha
{\gamma^i_j}^{(1)} {\gamma^j_p}^{(1)}C(R)^p_i +\frac{6}{r}\alpha^2
{\gamma^i_j}^{(1)}C(R)^j_pC(R)^p_i, \label{gamma}
\end{eqnarray}
and the anomalous dimension of the matter field calculated up to
two loops
\begin{eqnarray}
 {\gamma^i_j}&=& \frac 12 y^{ikl}y_{jkl}-2\alpha C(R)^i_j  \label{gamma1} \\
&-&(y^{imp}y_{jmn}+2\alpha C(R)^p_j\delta^i_n)(\frac 12
y^{nkl}y_{pkl}-2\alpha C(R)^n_p)+2\alpha^2QC(R)^i_j \ , \nonumber
\end{eqnarray}
one can get the solution
\begin{eqnarray}
{\Sigma_{\alpha}}^{(1)}&=&M^2\label{sigma1}, \\
 {\Sigma_{\alpha}}^{(2)}&=& \Delta_{\alpha}^{(2)}=
-2\alpha [ \frac{1}{r} {(m^2)}_j ^i {C(R)}_i ^j - M^2 C(G)],
\label{sigma2} \\
 {\Sigma_{\alpha}}^{(3)}&=&\Delta_{\alpha}^{(3)}={\displaystyle \frac{\alpha}{2r}}
[{\displaystyle \frac{1}{2}} (m^2)^i_n y^{nkl}y_{jkl}+
{\displaystyle \frac{1}{2}} (m^2)^n_j y^{ikl}y_{nkl}+ 2(m^2)^m_n
y^{ikn}y_{jkm} \nonumber \\ &&\hspace*{1cm}+  A^{ikl}A_{jkl}
-8\alpha M^2 C(R)^i_j ]C(R)_i^j -2\alpha ^2QC(G)M^2 \nonumber
\\ &&\hspace*{1cm} - 4\alpha ^2C(G) [ {\displaystyle
\frac{1}{r}} {(m^2)}^i_j {C(R)}^j_i - M^2 C(G)]. \label{sigma3}
\end{eqnarray}
These expressions for $\Delta_\alpha$ (\ref{sigma2},\ref{sigma3})
coincide with those obtained in Ref.~\cite{JJ9803} for the mass of
the $\epsilon$-scalars.

The nonzero $\Delta$-term modifies the expression for the $\beta$ function of
the  soft scalar mass  starting from the second loop. Substituting
eq.(\ref{sigma2}) into the expression for the differential operator $D_2$ gives
in two loops
\begin{eqnarray}
\left[\beta_{m^2}\right]^{i\ (2)}_j
&=&-(A^{ikp}A_{jkn}+\frac{1}{2}(m^2)^i_ly^{lkp}y_{jkn}
+\frac{1}{2}y^{ikp}y_{lkn}(m^2)^l_j
+\frac{2}{2}y^{ilp}(m^2)^s_ly_{jsn}\nonumber \\
&+&\!\frac{1}{2}y^{iks}(m^2)^p_sy_{jkn}+\! \frac{1}{2}y^{ikp}(m^2)^s_ny_{jks}
+4\alpha m_A^2C(R)^p_j\delta^i_n)(\frac{1}{2}y^{nst}y_{pst}-2\alpha C(R)^n_p)
\nonumber
\\ &-&(y^{ikp}y_{jkn}+2\alpha C(R)^p_j\delta^i_n)(\frac{1}{2}A^{nst}A_{pst}
+\frac{1}{4}(m^2)^n_ly^{lst}y_{pst} +\frac{1}{4}y^{nst}y_{lst}(m^2)^l_p
\nonumber \\&+&\frac{4}{4}y^{nlt}(m^2)^s_ly_{pst} -4\alpha m_A^2C(R)^n_p )  +
12\alpha^2m_A^2QC(R)^i_j \nonumber \\ &-&(A^{ikp}y_{jkn}-2\alpha
m_AC(R)^p_j\delta^i_n) (\frac{1}{2}y^{nst}A_{pst}+2\alpha m_AC(R)^n_p)
\nonumber \\ &-&(y^{ikp}A_{jkn}-2\alpha m_AC(R)^p_j\delta^i_n)
(\frac{1}{2}A^{nst}y_{pst}+2\alpha m_AC(R)^n_p) \nonumber \\ &+&4\alpha^2
C(R)^i_j [ \frac{1}{r} {(m^2)}^k_l {C(R)}^l_k - M^2 C(G)],
\end{eqnarray}
where  the last term is an extra contribution due to nonzero $\Delta_\alpha$.

To argue that a solution for $\Delta_\alpha$ exists in all orders of PT, one
can consider the so-called NSVZ-scheme \cite{NSVZ} where the anomalous
dimension $\gamma_\alpha$  is equal to
\begin{equation}\label{NSVZ}
  \gamma_\alpha^{NSVZ}= \alpha \frac{\displaystyle Q-2r^{-1}
   {\rm Tr}[\gamma C(R)]}{\displaystyle 1-2C(G)\alpha}.
\end{equation}
Then the solution for  $\Delta_\alpha$ to all orders is
\begin{equation}
  \Delta_\alpha^{NSVZ}=-2 \alpha \frac{\displaystyle
   r^{-1} {\rm Tr}[m^2 C(R)]-M^2C(G)}{\displaystyle 1-2C(G)\alpha}.
\end{equation}
It coincides with $X$ of Ref.~\cite{JJ9803}.

This problem has been also addressed in Ref.~\cite{Y}, where
originally the additional contribution to the soft term $\beta$
function was absent. In a comment to  paper \cite{Y} it is
suggested that the discrepancy can be eliminated by introducing
the term proportional to the mass of the $\epsilon$-scalar in the
superfield formalism
\begin{eqnarray}
\frac{\tilde{m}_A^2}{2}V^A_{\mu}V^A_{\nu}{\hat{g}}^{\mu\nu}=
\frac{\tilde{m}_A^2}{2}\int d^4\theta\bar{\eta}\eta
\frac{1}{16g^2}\bar{\sigma}_{\mu}^{\dot{\alpha}\alpha}\bar{D}_{\dot{\alpha}}
(e^{-2gV}D_{\alpha}e^{2gV})
\bar{\sigma}_{\nu}^{\dot{\beta}\beta}\bar{D}_{\dot{\beta}}
(e^{-2gV}D_{\beta}e^{2gV}) {\hat{g}}^{\mu\nu},\label{yesm}
\end{eqnarray}
where  ${\hat{g}}^{\mu\nu}$ is a $2\epsilon$-dimensional metric tensor.

Similar things were done in Ref.~\cite{AGLR}, where the appearance
an extra term in RGE for the soft scalar masses is due to additional
"evanescent" operator \cite{GMZ} in DRED scheme as in  eq.(\ref{yesm}).
It leads to additional contribution in higher loops.

However, whenever it is true, technically, it is complicated.  We propose here
the other solution of this problem.

\section{Illustration}

As an illustration of the described procedure, we consider the case of the
MSSM. Here instead of one there are three gauge couplings, and though the
recipe is still the same, one faces some problem of the general nature. We
obtain below the explicit solutions for the $\Sigma_{\alpha_i}$ terms that can
be of interest for the applications in higher loops.

In the MSSM we have three gauge and three Yukawa couplings and, to simplify the
formulas, we use the following notation $$\alpha_i \equiv
\frac{g_i^2}{16\pi^2}, \ \ i=1,2,3;\  \ \ Y_k \equiv \frac{y_k^2}{16\pi^2}, \ \
k=t,b,\tau.$$
 Then, the modified couplings (\ref{g}-\ref{y2}) take the form
 {\begin{eqnarray} \tilde{\alpha}_i&=&\alpha_i\left(1+M_i \eta+\bar
M_i \bar{\eta}+(M_i\bar M_i + \Sigma_{\alpha_i})\eta \bar{\eta}\right),
\label{ga}\\ \tilde{Y}_k&=&Y_k\left(1-A_k \eta -\bar A_k \bar{\eta}+ (A_k\bar
A_k+\Sigma_k) \eta \bar \eta\right),\label{ya}
\end{eqnarray}}
where $\Sigma_k$ is the sum of the soft masses squared corresponding to a given
Yukawa vertex
 $$\Sigma_t=\tilde m^2_Q+\tilde m^2_U+ m^2_{H_2}, \ \
\Sigma_b=\tilde m^2_Q+\tilde m^2_D+ m^2_{H_1}, \ \ \Sigma_\tau=\tilde
m^2_L+\tilde m^2_E+ m^2_{H_1}. $$
 Now the RG equations for a rigid theory can be written in a
universal form
\begin{equation}
\dot  a_i= a_i\gamma_i(a), \ \ \ \   a_i = \{\alpha_i, Y_k\}, \label{RG}
\end{equation}
where $\gamma_i(a)$ stands for a sum of  corresponding anomalous dimensions. In
the same notation, the soft terms (\ref{ga},\ref{ya}) take the form
\begin{equation}
\tilde{a}_i=a_i(1+m_i \eta+\bar m_i \bar{\eta}+S_i\eta \bar{\eta}),
\label{usoft}
\end{equation}
where  $m_i=\{M_i, -A_k\}$ and  $S_i=\{ M_i\bar M_i+\Sigma_{\alpha_i}, A_k\bar
A_k+\Sigma_k \}$.

 Substituting  eq.(\ref{usoft}) into eq.(\ref{RG}) and
expanding over $\eta $ and $\bar \eta$, one can get the RG equations for the
soft terms
\begin{equation}
\dot{\tilde{a}}_i= \tilde{a}_i\gamma_i(\tilde{a}). \label{RGs}
\end{equation}
Consider first the F-terms.  Expanding over $\eta$,  one has
\begin{equation}
 \dot m_i =  \left.\gamma_i(\tilde{a})\right|_F = \sum_j
a_j\frac{\partial \gamma_i}{\partial a_j}m_j. \label{rgem}
\end{equation} This is just the  RG equation for the soft terms $M_i$
and $A_k$~\cite{JJ9709,AKK}. Proceeding in the same way for the D-terms, one
gets after some algebra
\begin{equation} \dot S_i = 2m_i\sum_j
 a_j\frac{\partial \gamma_i}{\partial a_j}m_j+\sum_ja_j\frac{\partial
\gamma_i}{\partial a_j}S_j + \sum_{j,k} a_ja_k\frac{\partial^2
   \gamma_i}{\partial a_j  \partial a_k}m_jm_k.
\end{equation}
Substituting $S_i= m_i\bar m_i+\Sigma_i$, one has the following RG equation for
the mass terms
\begin{equation}
 \dot{\Sigma}_i =  \gamma_i(\tilde{a})\vert_D =
 \sum_ja_j\frac{\partial \gamma_i}{\partial a_j}(m_jm_j+\Sigma_j)  +
   \sum_{j,k} a_ja_k\frac{\partial^2 \gamma_i}{\partial a_j  \partial
a_k}m_jm_k.\label{rges}
 \end{equation}

Using the explicit form of  anomalous dimensions calculated up to some order,
one can reproduce the desired RG equations for the soft terms. In case of
squark and slepton masses, they contain the contributions from unphysical
masses $\Sigma_{\alpha_i}$. To eliminate them, one has to  solve the equation
for $\Sigma_{\alpha_i}$. In the case of the MSSM up to three loops, the
solutions are
\begin{eqnarray}
  {\Sigma}_{\alpha_1}&=&M_1^2-\alpha_1 \sigma_1 -\frac{199}{25}{\alpha}_1^2
M_1^2-\frac{27}{5}\alpha_1\alpha_2M_2^2
-\frac{88}{5}\alpha_1\alpha_3M_3^2\nonumber \\
&+&\frac{13}{5}\alpha_1Y_t(\Sigma_t+A_t^2)+\frac{7}{5}\alpha_1Y_b(\Sigma_b+A_b^2)
+\frac{9}{5}\alpha_1Y_{\tau}(\Sigma_{\tau}+A_{\tau}^2),\\
  {\Sigma}_{\alpha_2}&=&M_2^2-\alpha_2 (\sigma_2-4M_2^2) -{\alpha}_2^2(
4\sigma_2+9M_2^2)-\frac{9}{5}\alpha_2\alpha_1M_1^2
-24\alpha_2\alpha_3M_3^2\nonumber \\
&+&3\alpha_2Y_t(\Sigma_t+A_t^2)+3\alpha_2Y_b(\Sigma_b+A_b^2)
+\alpha_2Y_{\tau}(\Sigma_{\tau}+A_{\tau}^2),\\
{\Sigma}_{\alpha_3}&=&M_3^2-\alpha_3 (\sigma_3-6M_3^2) -{\alpha}_3^2(
6\sigma_3-22M_3^2)
-\frac{11}{5}\alpha_3\alpha_1M_1^2-9\alpha_3\alpha_2M_2^2\nonumber \\
&+&2\alpha_3Y_t(\Sigma_t+A_t^2)+2\alpha_3Y_b(\Sigma_b+A_b^2),
\end{eqnarray}
where we have used the combinations~\cite{MV}
\begin{eqnarray}
\sigma_1 &=&\frac 15 \left[3 ( m^2_{H_1} +   m^2_{H_2})+ 3(
 \tilde m^2_{Q}+3 \tilde m^2_{L} +8 \tilde m^2_{U}+2 \tilde m^2_{D}+6
 \tilde m^2_{E})\right],\label{s1}\\
\sigma_2 &=& m^2_{H_1} +   m^2_{H_2}+ 3(
 3\tilde m^2_{Q}+ \tilde m^2_{L}),\label{s2}\\
\sigma_3 &=& 3(
 2\tilde m^2_{Q}+ \tilde m^2_{U}+ \tilde m^2_{D}).\label{s3}
\end{eqnarray}

Notice, however, that the solutions (\ref{s1}-\ref{s3}) correspond to
particular boundary conditions, while, in general, one can use arbitrary ones.
Here we encounter the general problem that the solutions for  physical masses
depend on the unphysical parameter ($\epsilon$-scalar mass in the component
approach in the DRED scheme and an auxiliary field mass $\Delta$ in the
superfield approach).

 The solution to this paradox, mentioned also in
Ref.~\cite{JJMVY}, follows from the observation that the running soft masses
that obey the RG equations are not, strictly speaking, the observables and are
scheme-dependent. More appropriate are the pole masses, that are
scheme-independent. The authors of Ref.~\cite{JJMVY} proposed the solution of
the paradox by passing to the DRED$'$ scheme via the shift of the running soft
mass, which allows one to get rid of the unwanted $\epsilon$-scalar mass and
does not influence the pole mass. In one-loop order, the shift is
\begin{equation}
(m^2)_i^j|_{{\overline {DR}}\;{}^{'}}=(m^2)_i^j|_{{\overline
{DR}}}-\frac{2 g_A^2C_A(i)}{(4\pi)^2}\delta_i^j
\tilde{m}_\epsilon^2 ,
\end{equation}
where $\tilde{m}_\epsilon$ is the $\epsilon$-scalar mass. The procedure can be
continued in the same way in higher loops.

One can easily see how a similar trick works in our approach in case of one
gauge coupling (and, consequently, one $\Delta$ term). Indeed, consider
eq.(\ref{rges}). It is a linear inhomogeneous differential equation. Hence, to
any given solution of this equation one can add an arbitrary solution of a
homogeneous equation. In our case, the solution of a homogeneous equation is
\begin{equation}\label{h}
  \Sigma_i={\cal C}\gamma_i, \ \ \ i =\alpha_1,\alpha_2,\alpha_3,t,b,\tau,
\end{equation}
where ${\cal C}$ is an overall constant.

Hence, if one has the only gauge coupling one can choose the constant ${\cal
C}$ so that one can get any desirable boundary condition for $\Sigma_\alpha$.
The price for this is  extra terms in the other $\Sigma$'s (and soft masses)
proportional to the corresponding anomalous dimensions. However, the shift of
the running mass  by a term proportional to the anomalous dimension does not
change the pole mass, since it can be absorbed into the scale redefinition.
This is due to the fact that the coefficient  of the $\log\mu^2$ term is just
the anomalous dimension of the field.

Thus, the arbitrariness in the unphysical mass boundary condition does not
influence the physical masses.

However, one has only one overall constant ${\cal C}$, and the above argument
clearly works when one has only one gauge coupling. In case of many couplings,
it is more tricky, and we have not found an obvious explanation.

\section{Conclusion}

Summarizing, we would like to stress once again that soft breaking of
supersymmetry can be realized via interaction with an external superfield that
develops nonzero v.e.v.'s for its $F$ and $D$ components. In the superfield
notation, it can be reformulated as a modification of the rigid couplings that
become external superfields. The same is true for the gauge-fixing parameter
that can also be considered as a rigid coupling. The soft masses of scalar
particles obtain their contribution from the D-components of external
superfields. The latter  also lead to nonzero masses for unphysical degrees of
freedom, ghost and gauge auxiliary fields. These unphysical masses enter into
the RG equations for the physical scalars and have to be eliminated. This
creates an ambiguity in the running scalar masses; it can be resolved by
passing to the pole masses.

 \vglue 0.5cm
 {\bf Acknowledgments}

\vglue 0.5cm The authors are grateful to A.Bakulev, S.Mikhailov, G.Moultaka,
A.Onishchenko and A.Slavnov for useful discussions. Financial support from RFBR
(grants \# 99-02-16650 and 00-15-96691) is kindly acknowledged.

\section*{Appendix A}
\setcounter{equation}0
\renewcommand{\theequation}{A.\arabic{equation}}

The RG equation for the parameter $\xi$ in a rigid theory is
\begin{equation}\label{rxi}
  \dot \xi = - \gamma_V \xi,
\end{equation}
where $\gamma_V$ is the anomalous dimension of the gauge
superfield. To find the soft terms $x,\bar x$ and $z$, one should
solve the modified equation
\begin{equation}\label{rxim}
  \dot {\tilde \xi} = - \gamma_V(\tilde \alpha, \tilde y, \tilde \xi) \tilde
  \xi.
\end{equation}

In one-loop order  $\gamma_V= (b_1+b_2\xi) \alpha$, where $b_1+b_2=Q$,  and the
solutions are
\begin{eqnarray}
   x&=& -(M+x_0)\frac{b_1+b_2\xi}{Q}\ , \ \ \ \ \bar x =  -(\bar M +\bar
   x_0)\frac{b_1+b_2\xi}{Q}\ , \label{sol2} \\
    z &=& -(\Sigma_\alpha+z_0)\frac{b_1+b_2\xi}{Q} + \frac{b_2\xi}{Q}(M+x_0)(\bar M
    +\bar x_0)\frac{b_1+b_2\xi}{Q}\ ,
\end{eqnarray}
where  $x_0, \bar x_0$, and $z_0$ are arbitrary constants. In the Abelian case
when $b_1=Q, \ b_2=0$, the solutions are simplified and  can be chosen as
$$x=-M(1-\xi),\ \ \ \bar x = -\bar M (1-\xi), \ \ \  z=-\Sigma_\alpha (1-\xi)-
M\bar M \xi (1-\xi).$$ Together with the expression for $\tilde \alpha$
(\ref{Tildeg}) it gives eq.(\ref{ghs2}) above.

\end{document}